\documentclass[prb]{revtex4}
\usepackage[dvips]{graphics}

\usepackage{amsmath}

\begin{document}

\title{The radiation of a uniformly accelerated 
charge is beyond the horizon: A simple
derivation}

\author{Camila de Almeida}
\email{camila@fma.if.usp.br}
\affiliation{
Instituto de F\'\i sica, Universidade de S\~ao Paulo,
CP 66318, 05315-970 S\~ao Paulo, SP, Brazil}

\author{Alberto Saa}
\email{asaa@ime.unicamp.br}
\affiliation{Departamento de Matem\'atica Aplicada,
IMECC -- UNICAMP,
C.P. 6065, 13083-859 Campinas, SP, Brazil}


\begin{abstract}
By exploring some elementary consequences of the covariance of Maxwell's
equations under general coordinate transformations, we show that even though
inertial observers can detect electromagnetic radiation emitted from a
uniformly accelerated charge, comoving observers will see only a static
electric field. This analysis can add insight into one of the most
celebrated paradoxes of the last century.
\end{abstract}

\maketitle
\section{Introduction}

Paradoxes provide good opportunities to learn and teach physics. The
long-standing paradox about the electromagnetic radiation emitted by a
uniformly accelerated charge has received considerable attention. Eminent
figures such as Pauli, Born, Sommerfeld, Schott, von Laue, and many others
have contributed to this debate with different answers.\cite{EG} 
The relevant questions we consider are: Does a uniformly
accelerated charge actually radiate? In a constant gravitational field should
free-falling observers detect any radiation emitted by free-falling charges?
Is the equivalence principle valid for such situations?

If the answer to the first question is affirmative, a free-falling charge
will radiate according to an observer at rest, because in a constant
gravitational field, any particle should move with uniform acceleration.
However, an observer falling freely with the charge would observe it at
rest and  no radiation at all. How can this answer be compatible with
an affirmative answer to the first question? Moreover, if the equivalence
principle is assumed to be valid, we would conclude that a charged particle
at rest on a table should radiate, because for free-falling inertial
observers the particle is accelerating. To explain this puzzle, we need to
recognize that the concept of radiation has no absolute meaning, and that it
depends both on the radiation field and the state of motion of the
observer.\cite{Rohrlich,R2,Boulware,Peirls} This dependence is the main
conclusion of a celebrated and long debate, exhaustively presented in the
recent series of papers by Eriksen and Gr\o n,\cite{EG} where the
reader can find relevant references.

Inertial observers have no doubts about the answer to the first question.
They will answer it affirmatively by using special relativity and Maxwell's
equations, as it is done in classical electrodynamics texts (see, for
example, Ref.~\onlinecite{J} whose conventions we adopt). Nevertheless,
comoving observers, that is, accelerated observers with respect to whom the
charge is at rest, will not detect any radiation because the radiation field
is confined to a spacetime region beyond a horizon that they cannot
access.\cite{Boulware} As we will see, uniformly accelerated observers are
able, in principle, to detect electromagnetic radiation from an
inertial charge. These observations are enough to solve the paradoxes posed
by the three questions. The last two conclusions are obtained by considering
Maxwell's equations for noninertial observers. We will show that we can
conclude that comoving observers have no access to the radiation field of a
uniformly accelerated charge. The concept of a horizon emerges naturally in 
this context. Our approach is inspired by the recent
analysis of this problem by Gupta and Padmanabhan\cite{GP} whose result is
much stronger. They show how to calculate the radiation field for an
arbitrarily moving charge by solving Maxwell's equations in the
(noninertial) reference frame where the charge is at rest, and then
transforming the electromagnetic field to the inertial reference frame by
exploiting the tensorial character of Maxwell's equations. As stressed in
Ref.~\onlinecite{GP}, this is a   remarkable result because the spatial and time
dependences of the radiation field in the inertial frame have to be
converted into the geometrical properties of the background metric of the
reference frame where the charge is static. We take the converse approach
here and will show that it is simple to conclude that the electromagnetic
field generated by a uniformly accelerated charge is observed by a comoving
observer as a purely electrostatic field.

\section{The hyperbolic motion}
The speed of light $c$ is the maximum speed that a physical body can attain.
Thus the uniformly accelerated motion of a particle should have
$|v|\rightarrow c$ as
$\tau\rightarrow\pm\infty$, where $\tau$ is the proper time as measured by a
comoving clock.\cite{J} It is easy to deduce that a particle moving with
constant proper acceleration $g$ along the $z$ direction has a hyperbolic
worldline given by the curve $r^a(\tau)$:
\begin{subequations}
\label{hyper}
\begin{align}
ct &= r^0(\tau) =\frac{c^2}{g}\sinh \frac{g\tau}{c}, \\
x &= r^1(\tau) = 0,\\
y &= r^2(\tau) = 0, \\
z &= r^3(\tau) =\frac{c^2}{g}\cosh \frac{g\tau}{c}.
\end{align}
\end{subequations}
There is no loss of generality if the motion is
restricted to the $z$ direction. Such a worldline is displayed in
Fig.~\ref{fig1}.

\subsection{The Horizons}

The velocity of a particle according to Eq.~(\ref{hyper})
approaches $\pm c$ as 
$\tau \rightarrow\pm \infty$, and
its trajectory tends asymptotically to the lines $\pm ct=z$, with $z>0$ as
shown in Fig.~\ref{fig1}. Consider the point $Q$. Its past light-cone
intersects the hyperbolic trajectory. Indeed, a large part of the trajectory
($\tau <
\tau^Q_{\rm ret}$)  is entirely
contained inside its past light-cone, implying that the point $Q$ could be
causally influenced by signals emitted by the particle for $\tau \le
\tau^Q_{\rm ret}$; no signal emitted for $\tau > \tau^Q_{\rm ret}$ will
influence $Q$, because the points of the trajectory with $\tau >
\tau^Q_{\rm ret}$ are not contained in the past light-cone of $Q$. Moreover,
a signal emitted in the space time $Q$ will affect only the region
corresponding to its future light-cone, implying that no signal emitted in
$Q$ will reach the particle moving according to Eq.~\eqref{hyper}. The line
$ct=z$ acts as a {\em future event horizon} for regions I and IV, or,
equivalently, a {\em past event horizon} for regions II and III. No signal
emitted in regions II or III will reach regions I and IV, although signals
emitted in I or IV can cross the line and enter into regions II and III.
Analogously, the line
$-ct=z$ is a future event horizon for regions III and IV, or a past event
horizon for I and II.

Because the hyperbolic trajectory is entirely contained in region I, the
lines
$ct=z$ and $-ct=z$ act, respectively, as the future and past horizons for a
particle under uniformly accelerated motion. We will see that such
structures appear naturally when we consider the radiation emitted by a
uniformly accelerated charged particle.

\subsection{The Radiation in the inertial frame}

\begin{figure}[ht]
\resizebox{0.4\linewidth}{!}{\includegraphics*{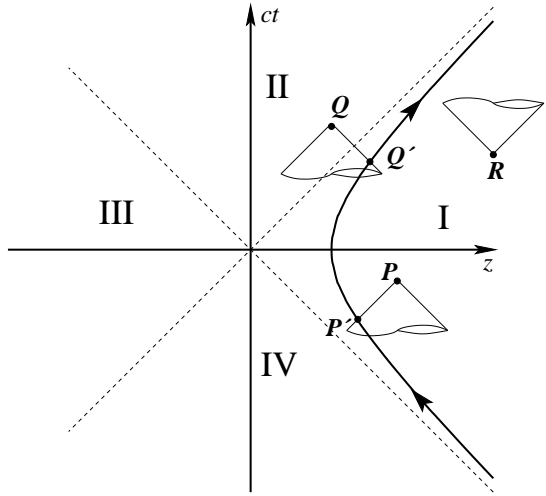}}
\caption{The hyperbolic trajectory $r^a(\tau)$ given by Eq.~(\ref{hyper}).
The retarded time $\tau_{\rm ret}$ associated with a given point
$(ct,x,y,z)$ corresponds to the (unique) intersection of
the past light-cone of $(ct,x,y,z)$ with the trajectory $r^a(\tau)$. For
instance, 
$Q'=(c^2/g)(\sinh(g\tau_{\rm
ret}^Q/c),0,0,\cosh(g\tau_{\rm ret}^Q/c))$ and 
$P'=(c^2/g)(\sinh(g\tau_{\rm
ret}^P/c),0,0,\cosh(g\tau_{\rm ret}^P/c))$ define,
respectively, the retarded times $\tau_{\rm ret}^Q$ and $\tau_{\rm ret}^P$
associated with the points
$Q$ and $P$. The future light-cone is the boundary of the causal future of
a given point. Thus, any event occurring, for instance, in the spacetime
point $R$ will affect only the region enclosed by its future light-cone,
with the light-cone surface reserved only to
signals moving with velocity $c$.
Note that only regions I and II are affected by the fields due to
a charged particle with a worldline given by Eq.~(\ref{hyper}).}
\label{fig1}
\end{figure}

The metric of the inertial Minkowski spacetime is given by
\begin{equation}
ds^2 = \eta_{ab}dx^a dx^b= c^2dt^2 - dx^2 - dy^2 - dz^2,
\end{equation}
where $x^a=(ct,x,y,z)$.
Maxwell's equations are not only Lorentz invariant, they can also be cast
in a generally covariant way, valid for any reference frame with the metric
$G_{ab}$,
\begin{equation}
\label{hom}
\partial_a F_{bc} + \partial_b F_{ca} + \partial_c F_{ab} = 0,
\end{equation}
\begin{equation}
\label{nhom}
\frac{1}{\sqrt{G}} \partial_a ( \sqrt{G} F^{ab}) = \frac{4\pi}{c}
J^b,
\end{equation}
where $G=|\det G_{ab}|$ and $J^{b}$ is the external
4-current.\cite{J,mould} Equation~(\ref{hom}) is automatically satisfied if
the 4-potential $A_b$ is introduced: $F_{bc} = \partial_b A_c - \partial_c
A_b$. In the inertial Minkowski frame, the radiation emitted by a uniformly
accelerated charge $e$ corresponds to the solution of Eqs.~(\ref{hom}) and
(\ref{nhom}) with $G_{ab}=\eta_{ab}$, 
$J^a(x)=ec\int d\tau V^a(\tau)\delta^{(4)}(x-r(\tau))$, and $V^a =
\dot{r}^a$. Such a solution is given by:\cite{J}
\begin{equation}
\label{sol}
F^{ab} = e\Big[ \frac{1}{V^c (x_c-r_c)} \frac{d}{d\tau} 
\frac{(x^a-r^a)V^b - (x^b-r^b) V^a}{V^c (x_c-r_c)}
\Big]_{\rm ret},
\end{equation}
where the quantity between the brackets is to be evaluated at the retarded
time
$\tau_{\rm ret}$ given by (see Fig.~\ref{fig1}): 
\begin{equation}
[x^a-r^a(\tau_{\rm ret})][x_a-r_a(\tau_{\rm ret}) ]= 
\Big(ct-\frac{c^2}{g}\sinh \frac{g\tau_{\rm ret}}{c} \Big)^2 - \rho^2 - 
\Big( z-\frac{c^2}{g}\cosh \frac{g\tau_{\rm ret}}{c} \Big)^2 =0,
\end{equation}
with $\rho^2=x^2+y^2$,
leading to
\begin{equation}
\label{ret}
z\cosh \frac{g\tau_{\rm ret}}{c} - ct\sinh \frac{g\tau_{\rm ret}}{c} = 
\frac{g}{2} \Big(\frac{\rho^2}{c^2} + \frac{z^2}{c^2} - t^2 +\frac{c^2}{g^2} 
\Big).
\end{equation}

In the inertial frame we can read from $F^{ab}$
the usual three-dimensional components of the electric and
magnetic field as
\begin{equation}
F^{ab} = 
\begin{pmatrix} 
0 & -E_x & -E_y & -E_z \\
E_x & 0 & -B_z & B_y \\
E_y & B_z & 0 & -B_x \\
E_z & -B_y & B_x & 0
\end{pmatrix} .
\end{equation}
If we use the equation $V^a(x_a - r_a) = c(ct\cosh \frac{g\tau}{c} -
z\sinh \frac{g\tau}{c})$, we obtain after some straightforward
algebra,
\begin{subequations}
\label{Ex}
\begin{align}
\frac{1}{xz}E_x & = \frac{1}{yz}E_y = 
\frac{4}{g^2(\frac{\rho^2}{c^2} + \frac{z^2}{c^2} -
t^2)^2-\frac{c^4}{g^2}}E_z \\ 
\frac{1}{ctx}B_y & = -\frac{1}{cty}B_x= 
\frac{eg}{c^2\Big(ct\cosh \frac{g\tau_{\rm ret}}{c} - z\sinh
\frac{g\tau_{\rm ret}}{c}\Big)^3 }, 
\end{align}
\end{subequations}
and $B_z = 0$,
where Eq.~(\ref{ret}) was explicitly used in the expression for $E_z$.
These are the electromagnetic fields due to a uniformly accelerated charge
moving according to Eq.~(\ref{hyper}).

The radiation content can be extracted by separating the components that
drop off as $1/R$ from the usual Coulomb $1/R^2$ fields.\cite{J} As shown in
Fig.~\ref{fig1}, only regions I and II can experience the fields in
Eq.~(\ref{Ex}). The main conclusion of Ref.~\onlinecite{Boulware} is that,
even though radiation components are present in both regions, only
observations performed in region II (and, perhaps, also on the boundary
$ct=z$ between regions I and II) would allow us to detect unambiguously the
radiation emitted by the charge. This conclusion implies that the comoving
observer would not detect any radiation at all, because region II is
inaccessible to uniformly accelerated observers.\cite{Boulware} Although
this conclusion is correct, its logical derivation is involved and not
intuitive. Two regions of spacetime which have qualitatively distinct
behavior for the radiation field according to inertial observers are
identified and, then, it is shown  that the comoving
observers have access only to the region where inertial observers are not
able to detect any radiation field.\cite{Boulware}  However, this conclusion does not
directly imply that the comoving observers cannot detect the radiation
because, as we have discussed, the detection of radiation has no absolute
meaning because the detection depends both on the radiation field and the
state of motion of the observer.

\subsection{No Radiation in the comoving frame}

We can show directly that a comoving observer will observe the fields in
Eq.~(\ref{Ex}) as a static electric field. The reference frame of a uniformly accelerated
observer corresponds to the Rindler spacetime,\cite{mould} which in our
case is spanned by the coordinates ${x'}^a(x)=(c\tau(t,z),x,y,\xi(t,z))$
defined by
\begin{subequations}
\label{rindler}
\begin{align}
t &= \sqrt{\frac{2\xi}{g}}\sinh \frac{ g\tau}{c} \\
z &= c\sqrt{\frac{2\xi}{g}}\cosh \frac{ g\tau}{c},
\end{align}
\end{subequations}
with $\xi>0$. The particle under the hyperbolic motion (\ref{hyper}) in
the Rindler reference frame is at rest at $\xi=c^2/2g$,
and its proper time is measured by $\tau$. In these coordinates the
spacetime interval is given by
\begin{equation}
\label{rind}
ds^2 = G_{ab}dx^adx^b = 2g\xi d\tau^2 - dx^2 - dy^2-
c^2\frac{d\xi^2}{2g\xi}. 
\end{equation}
Note that the coordinates defined by Eq.~(\ref{rindler}) cover only region I
of the original Minkowski spacetime. Because static observers ($\xi$ is a
constant) correspond to uniformly accelerating observers in the original
Minkowski spacetime, their velocity in the inertial frame will approach
$c$ as $\tau\rightarrow\infty$, implying that no signal coming from region
II will ever reach them (see Fig.~\ref{fig1}). As mentioned, the
line
$ct=z$ behaves as an event horizon, and no signal emitted in regions II or
III can escape into regions I and IV.

The coordinate transformation (\ref{rindler}) can be used to obtain the
solution of Maxwell's equations (\ref{hom}) and (\ref{nhom}) for the
Rindler spacetime with a charge $e$ at rest in $\xi=c^2/2g$. Recall that the
electromagnetic field $F^{ab}$ is a tensor and hence under a coordinate
transformation $x^a \rightarrow {x'}^a(x)$ it transforms as\cite{J,mould}
\begin{equation}
\label{tra1}
{F'}^{ab} = \frac{\partial {x'}^a}{\partial x^c}\frac{\partial
{x'}^b}{\partial x^d}F^{cd}.
\end{equation}
Because Maxwell's equations (\ref{hom}) and (\ref{nhom}) are
covariant under general coordinate transformations, 
${F'}^{ab}$ will be a solution for the coordinate system ${x'}^a(x)$ if
$F^{ab}$ is a solution of Eqs.~(\ref{hom}) and (\ref{nhom}) in the
coordinate system $x^a$. The magnetic components of ${F'}^{ab}$ are given by
\begin{subequations}
\label{tra}
\begin{align}
{F'}^{13} &= \frac{1}{c}\frac{\partial \xi}{\partial t} E_x + \frac{\partial
\xi}{\partial z} B_y, \\
{F'}^{23} &= \frac{1}{c}\frac{\partial
\xi}{\partial t} E_y - \frac{\partial \xi}{\partial z} B_x, \\
{F'}^{12} &= B_z = 0. 
\end{align}
\end{subequations}
Strictly speaking, 
we need to be careful about the interpretation of ${F'}^{13}$, ${F'}^{23}$,
and 
${F'}^{12}$ as the components of the magnetic field as observed in the
Rindler reference frame. A proper definition of electric and magnetic fields
for noninertial reference frames can be obtained 
from the Lorentz force formula.\cite{GP} This issue will not be
relevant to our analysis.
 
By using Eq.~(\ref{Ex}), we have
\begin{equation}
\frac{1}{y}{F'}^{13} = \frac{1}{x}{F'}^{23} = 
\frac{eg}{c^2\Big(ct\cosh \frac{g\tau_{\rm ret}}{c} - z\sinh
\frac{g\tau_{\rm ret}}{c}\Big)^3}
\Big( \frac{1}{c}\frac{\partial \xi}{\partial t} z+c\frac{\partial
\xi}{\partial z} t\Big).
\end{equation}
From the transformation (\ref{rindler}), we can calculate
\begin{subequations} 
\label{rr}
\begin{align}
\frac{\partial \tau}{\partial t} &= \frac{z}{2\xi}, \qquad 
\frac{\partial \tau}{\partial z} = -\frac{t}{2\xi},& \\
\frac{\partial \xi}{\partial t} &= -{gt} , \qquad 
\frac{\partial \xi}{\partial z} = \frac{gz}{c^2},&
\end{align}
\end{subequations}
leading to ${F'}^{13}={F'}^{23}=0$. Therefore the only nonvanishing
components of the electromagnetic field experienced by comoving observers
are ${F'}^{01}$, ${F'}^{02}$, and ${F'}^{03}$. Because
the only
nonvanishing component of the 4-current ${J'}^a$ is ${J'}^0$ for a static
charge in the Rindler reference frame, we conclude from
Eq.~(\ref{nhom}) that the remaining nonvanishing components of the
electromagnetic field are static, that is, $\partial_0{F'}^{01} =
\partial_0{F'}^{02} = \partial_0{F'}^{03} = 0$, so that there is no
radiation field in region I, the Rindler reference frame.

This result answers our question. A comoving observer will not detect any
radiation from a uniformly accelerated charge. The comoving observer can
receive signals only from regions I and IV. The field emitted by the
accelerated charge does not reach region IV, and in region I, it is
interpreted by the comoving observer as a static field. We note that
essentially the same argument was used by Rohrlich to show that in a static
homogeneous gravitational field, static observers do not detect any
radiation from static charges.\cite{R2}

The situation is qualitatively different beyond the horizon in region II.
Although uniformly accelerated observers will never receive any information
from region II, they can affect this region.
The coordinate
system (\ref{rindler}) can be extended to include region II by considering
$\xi<0$ and
\begin{subequations}
\label{milne}
\begin{align}
t &= \sqrt{\frac{-2\xi}{g}}\cosh \frac{g\tau}{c} \\
z &= c\sqrt{\frac{-2\xi}{g}}\sinh \frac{g\tau}{c}.
\end{align}
\end{subequations}
The metric (\ref{rind}) and the expressions (\ref{rr}), valid 
for region I, also hold in region II, but
with a crucial difference due to the change of sign of the metric
components: in region II,
$\xi$ instead of $\tau$ plays the role of a time parameter. Thus the
metric (\ref{rind}) is not static in region II. The magnetic components of
${F'}^{ab}$ in region II can be obtained from transformations such as
Eq.~(\ref{tra}) if we take into account that the components 0
(temporal) and 3 (spatial) are, respectively, $\xi$ and $c\tau$:
\begin{subequations}
\label{F}
\begin{align}
\frac{1}{y}{F'}^{13} = \frac{1}{x}{F'}^{23} &= 
\frac{eg}{c^2\Big(ct\cosh \frac{g\tau_{\rm ret}}{c} - z\sinh
\frac{g\tau_{\rm ret}}{c}\Big)^3}
\Big( \frac{\partial \tau}{\partial t} z+c^2\frac{\partial \tau}{\partial
z} t\Big) \\
&= - \frac{e}{\Big(ct\cosh \frac{g\tau_{\rm
ret}}{c}- z\sinh \frac{g\tau_{\rm ret}}{c}\Big)^3}.
\end{align}
\end{subequations}
The fields (\ref{F}), together with the electric components that can be
obtained in an analogous way, are time-dependent solutions of the (vacuum)
Maxwell equations in region II, having radiating parts.\cite{note} However,
they are inaccessible to a comoving observer because they are confined
beyond his/her future horizon.

\section{Concluding Remarks}
\begin{figure}[ht]
\resizebox{0.4\linewidth}{!}{\includegraphics*{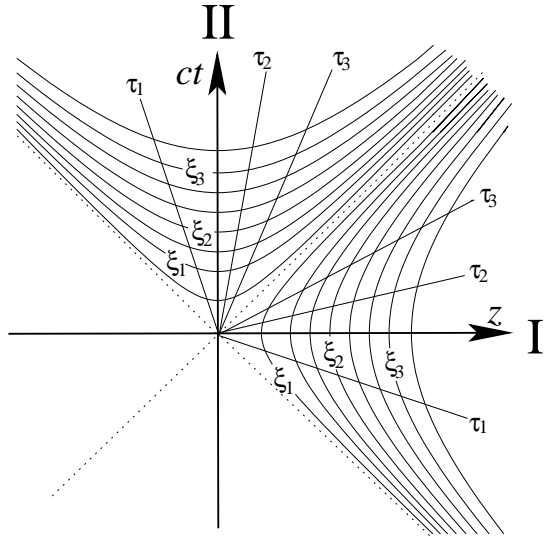}}
\caption{The lines of constant $\xi$ and $\tau$ according to
Eqs.~(\ref{rindler}) and (\ref{milne}), respectively, for the regions I and
II. In region I, the Rindler frame where $\xi>0$, the identified lines
correspond to
$\xi_1 < \xi_2 < \xi_3$ and 
$\tau_1 < \tau_2 < \tau_3$. Lines of constant $\xi$ (the hyperbola) are
timelike. On the other hand, for region II, known as the Milne frame
where $\xi<0$,
the lines of constant $\tau$ are timelike. The identified lines in II
correspond to the cases $\tau_1 < \tau_2 < \tau_3$ and $\xi_1 > \xi_2 >
\xi_3$. The horizon, the boundary
$ct=z$ between I and II, corresponds to one half of the degenerated 
hyperbola corresponding to
$\xi=0$.}
\label{fig2}
\end{figure}

The physics of the Rindler space is sufficiently subtle to deserve
some extra remarks. Trajectories with constant $\xi$ (see Fig.~\ref{fig2})
correspond to uniformly accelerated trajectories in the inertial frame, but
with {\em distinct} accelerations. The trajectory (\ref{hyper})
corresponds to the static worldline $\xi=c^2/2g$ in the Rindler frame. From
the inertial frame point of view, the true comoving observer should
correspond also to $\xi=c^2/2g$, because any other static Rindler observer
would be in relative motion according to the inertial frame point of view
with respect to the charge following Eq.~(\ref{hyper}). Our results
show that the electromagnetic field of the uniformly accelerated charge is
realized as a purely electrostatic field everywhere in the Rindler frame,
implying that even observers with $\xi\ne c^2/2g$, for which the
charge is indeed accelerating when observed from the inertial point of view, 
would not detect the emitted radiation.
This observation, which anticipates an intriguing quantum
result described by Matsas,\cite{accelob} 
reinforces the role played by the horizon, the unique  property
that the trajectories of these distinct observers have in common. (see Fig.~\ref{fig2})

The discussion can be considerably enriched by the introductions of 
quantum mechanical concepts. The classical radiation emitted by the
accelerated charge in the inertial frame consists of a large number of real
photons, which due to some subtle quantum effects cannot be detected by
comoving observers.\cite{Unruh} To illustrate the novelties brought 
by quantum mechanics, consider in the Minkowski space a uniformly
accelerated observer following a trajectory such as that in
Eq.~(\ref{hyper}) and a charge at rest at the origin. The worldline for this
charge is the $ct$ axis, and it is restricted to regions II and IV. The
solution of Maxwell equations in the inertial frame is the static Coulomb
field
\begin{equation}
\frac{1}{x}E_x = \frac{1}{y}E_y = \frac{1}{z}E_z =
\frac{e}{(x^2+y^2+z^2)^{3/2}},
\end{equation}
which spreads over all four regions of Fig.~\ref{fig1}. In region I, where a
uniformly accelerated observer can detect any field coming from the
charge, the static Coulomb field will be measured by such observers 
as a time-depending electromagnetic
field with components
\begin{align}
\label{el}
{F'}^{01} & = \frac{z}{2\xi} E_x, \qquad {F'}^{02} = 
\frac{z}{2\xi} E_y,\qquad {F'}^{03} = E_z \nonumber \\
{F'}^{31} &= -\frac{gt}{c} E_x, \qquad {F'}^{32} = -\frac{gt}{c} E_y,\qquad
{F'}^{12} = 0.
\end{align}
The fields (\ref{el}) are solutions of the (vacuum) Maxwell equations in
region I as seen by uniformly accelerated observers. These fields have
radiative components, although it is not so easy in this case to identify
the terms dropping off as $1/R$.\cite{note} Note that the accelerated
observers are completely unaware of the charge fate in region II. Because
they can detect only the contributions coming from regions I and IV, they
will never discover what eventually happens to the charge in region II if
it, for instance, accelerates or even if it vanishes.

An analysis of this problem based on quantum field theory, however,
demands that the trajectory of the
charge be entirely inside region I. Is it possible to
conclude something in this case? Astonishing the answer is
yes.\cite{accelob} In the Rindler reference frame static trajectories with
$\xi\rightarrow\infty$ (see Fig.~\ref{fig2}) correspond to uniformly
accelerated trajectories in the inertial frame, restricted by construction
to region I, but with proper acceleration $g=c^2/2\xi\rightarrow 0$. Thus,
they correspond to inertial trajectories! Now we can answer the question of
whether uniformly accelerated observers could detect any photon emitted by
these specific inertial charges, and the answer is no.\cite{accelob}

The detection of radiation is not the
only paradox involving accelerated charges. Another very interesting
paradox is related to the radiation reaction force. As we discussed, an
inertial observer detects the radiation emitted by a uniformly
accelerated charge. He/she can even calculate the (nonvanishing) total
radiated power. But we know from classical electrodynamics
that the radiation reaction force vanishes for a constant proper
acceleration.\cite{J} Hence, what is acting as the source of the radiated power? How
is it possible to conserve energy in this case? Interesting questions,
but that's another story \ldots\cite{f}

\acknowledgments

This work was supported by FAPESP and CNPq. The authors are grateful to
J.\ Casti\~neiras, G.\ E.\ A.\ Matsas and R.\ A.\ Mosna for enlightening
discussions. A.S.\ expresses his appreciation for the hospitality of 
the Abdus Salam International Centre for Theoretical Physics, Trieste,
Italy, where this work was partly done.

\newpage
\section*{Figure Captions}

\end{document}